\documentclass{article}

\usepackage[accepted]{icml2026}

\usepackage{microtype}
\usepackage{graphicx}
\usepackage{booktabs}
\usepackage{amsmath,amssymb,amsthm}
\usepackage{array}
\usepackage{enumitem}
\usepackage{threeparttable}
\usepackage{hyperref}

\graphicspath{{./}{figures/}}

\theoremstyle{definition}
\newtheorem{definition}{Definition}
\setlist{leftmargin=*,topsep=2pt,itemsep=1pt,parsep=0pt}

\newcommand{\widepaperfig}[3]{%
  \IfFileExists{#1}{%
    \includegraphics[width=#2,height=0.32\textheight,keepaspectratio]{#1}%
  }{%
    \fbox{\parbox[c][0.22\textheight][c]{#2}{\centering\small
    \textbf{Missing figure:} \texttt{#1}\\[2pt]#3}}%
  }%
}
\newcommand{\colpaperfig}[3]{%
  \IfFileExists{#1}{%
    \includegraphics[width=#2,height=0.22\textheight,keepaspectratio]{#1}%
  }{%
    \fbox{\parbox[c][0.16\textheight][c]{#2}{\centering\small
    \textbf{Missing figure:} \texttt{#1}\\[2pt]#3}}%
  }%
}

\icmltitlerunning{When Does Structure Help?}

\begin{document}

\twocolumn[
\icmltitle{When Does Structure Help?\\
The Information Bonus of AlphaFold2 Representations\\
over Protein Language Models}

\begin{icmlauthorlist}
\icmlauthor{Kargi Chauhan}{ucsc}
\end{icmlauthorlist}
\icmlaffiliation{ucsc}{University of California, Santa Cruz}
\icmlcorrespondingauthor{Kargi Chauhan}{kchauha3@ucsc.edu}
\icmlkeywords{AI for science, protein representation learning, AlphaFold2,
ESM-2, linear probing, allostery, molecular dynamics, benchmark leakage}
\vskip 0.25in
]

\printAffiliationsAndNotice{}

\begin{abstract}
AI scientist systems increasingly choose biological foundation models before
they choose experiments.  In protein pipelines, this creates a concrete
engineering and scientific question: \emph{when is the cost of structural
inference worth paying over a cheaper sequence-only model?}  We introduce the
\textbf{information bonus} (IB), a task-level metric that measures the
linearly accessible advantage of frozen single-sequence AlphaFold2 Evoformer
representations over frozen ESM-2 embeddings under protein-level
cross-validation.  Across
binding affinity regression (PDBbind, $n{=}5{,}680$), conformational
flexibility (ATLAS molecular dynamics, 268 proteins), and allosteric-site
classification (AlloSigDB, $n{=}9{,}925$ residues), IB is sharply
mechanism-dependent.  ESM-2 dominates binding affinity (IB$=-0.141$;
Pearson $r{=}0.449$ vs.\ $0.307$) and binary flexibility (IB$=-0.060$;
AUROC $0.824$ vs.\ $0.764$; $p{=}0.0017$).  AF2 single representations give
the only above-chance allostery predictions (IB$=+0.064$; AUROC $0.548$ vs.\
$0.485$), revealing long-range geometric signal not recovered from sequence
alone.  We also identify a residue-level leakage artifact: naive residue
splits inflate RMSF performance by 27--39\% depending on the representation,
enough to reverse representation rankings.  These results turn representation selection into a
measurable decision for AI-for-science systems.
\end{abstract}

\section{Introduction}
\label{sec:intro}

Protein foundation models are now infrastructure for AI-assisted discovery.
Sequence models such as ESM-2 can produce residue embeddings in seconds from a
single amino-acid string \citep{lin2023evolutionary}.  AlphaFold2 produces
geometry-aware Evoformer representations, but at the price of structural
inference, specialized software, and substantially higher compute
\citep{jumper2021highly,mirdita2022colabfold}.  For an autonomous or
semi-autonomous AI scientist, this choice is not cosmetic.  It determines
latency, cost, experimental throughput, and sometimes the scientific
conclusion itself \citep{lu2024aiscientist}.

The usual advice is that ``structure matters for structural tasks.''  That is true
but not operational.  Binding affinity, flexibility, and allostery are all
structural in some sense, yet they differ in the information that a model must
recover.  Binding sites are constrained by evolution.  Flexible regions often
leave sequence signatures of disorder, charge patterning, and low packing.
Allostery, by contrast, depends on non-local communication through the folded
contact network.  A useful benchmark should distinguish these regimes rather
than assume one representation is globally superior.

We propose the \textbf{information bonus} (IB): the held-out performance
difference between the best AF2 representation and ESM-2 on the same task,
using the same frozen linear probe and the same protein-level split.  IB$>0$
means structural pre-training adds usable signal; IB$<0$ means sequence
embeddings are sufficient or better.  This framing makes representation
choice measurable, cheap to estimate, and directly actionable inside
AI-for-science workflows.

\paragraph{Contributions.}
We make four contributions.  First, we define IB as a simple benchmark unit
for deciding when structural representations are worth their cost.  Second,
we show that ESM-2 is stronger for binding affinity and binary flexibility,
despite having no explicit coordinates.  Third, we identify allostery as the
clearest positive-IB regime, where AF2 is the only representation above
chance.  Fourth, we quantify a 27--39\% residue-level cross-validation inflation
effect in RMSF probing and show why protein-level GroupKFold is the correct
protocol for per-residue tasks.

\begin{figure*}[t]
\centering
\widepaperfig{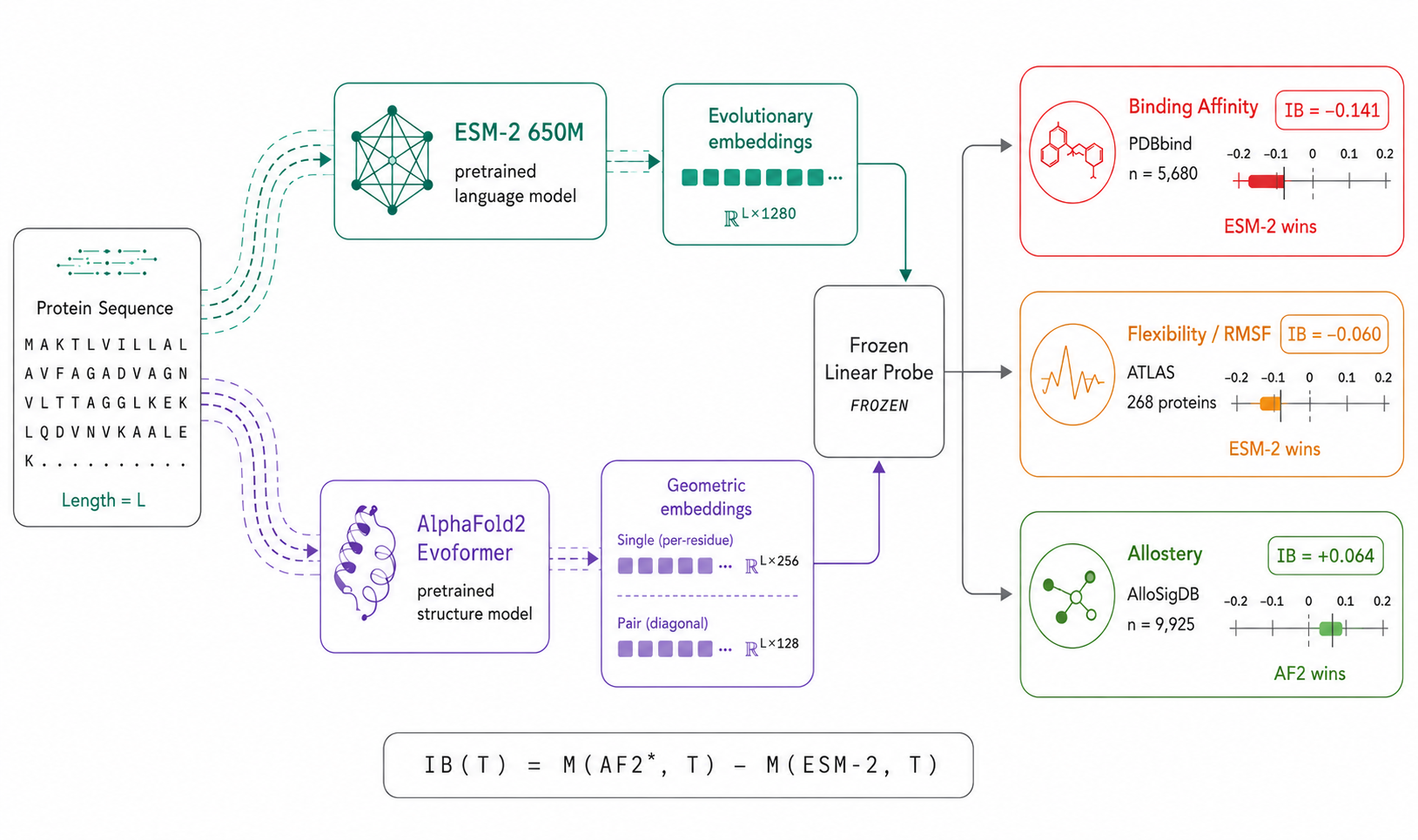}{0.92\textwidth}{Architecture diagram: sequence input,
ESM-2 path, AF2 Evoformer path, frozen probe, and task-specific IB outputs.}
\caption{\textbf{Information Bonus evaluation framework.}
Both models receive the same protein sequence.  ESM-2 supplies sequence-only
evolutionary embeddings; AlphaFold2 supplies Evoformer single and pair-diagonal
representations.  Frozen linear probes evaluate each representation on
binding, flexibility, and allostery.  The information bonus is the held-out
performance difference between the best AF2 representation and ESM-2.}
\label{fig:arch}
\end{figure*}

\section{Related Work}
\label{sec:related}

\paragraph{Protein representation learning.}
AlphaFold2 uses Evoformer blocks to iteratively update residue-level and
pairwise representations before coordinate prediction
\citep{jumper2021highly}.  ESM-1b, ProtTrans, and ESM-2 instead learn from
large protein sequence corpora and recover structural and functional
regularities without coordinate supervision
\citep{rives2021biological,elnaggar2021prottrans,lin2023evolutionary}.
Recent work has shown that sequence embeddings alone can match or exceed
structure-based methods on tasks such as virtual screening
\citep{lam2024virtualscreen} and binding affinity prediction
\citep{piao2025bindpred}.  Structure-aware language models such as SaProt
combine sequence and structural tokens \citep{su2023saprot}, underscoring the
open question studied here: which downstream tasks actually require explicit
geometric information?

\paragraph{Probing and interpretability.}
Linear probes measure information that is already linearly accessible in a
frozen representation \citep{alain2016understanding}.  TAPE established this
style of evaluation for protein transfer learning \citep{rao2019evaluating},
while later work showed that protein language models encode contacts,
mutation effects, and functional constraints
\citep{vig2021bertology,meier2021language,hie2022evolu}.  Recent sparse
autoencoder studies provide a complementary interpretability view: ESM-2
embeddings contain features aligned with domains, binding sites, structural
motifs, Gene Ontology annotations, and protein families
\citep{gujral2025sparse,simon2025interplm}.  Our work asks a different
question: when does AF2 add task-relevant information beyond those sequence
features?

\paragraph{Benchmarks and leakage.}
PDBbind provides measured protein-ligand affinities \citep{liu2017pdbbind};
ATLAS provides molecular-dynamics-derived flexibility labels
\citep{vander2024atlas}; and AlloSigMA/AlloSigDB organize experimentally
supported allosteric mechanisms \citep{li2022allosigdb}.  Molecular
benchmarks are vulnerable to scaffold, homolog, and identity leakage
\citep{yang2019analyzing,varadi2022alphafold}.  Grouped cross-validation and
careful split design are increasingly recognized as essential for protein
tasks \citep{fang2023imbalance,bushuiev2024leakage,bernett2024cracking}.  We
isolate a specific residue-level version of this problem: if residues from the
same protein appear in both train and test folds, probes can exploit protein
identity rather than transferable residue physics.

\section{Information Bonus}
\label{sec:ib}

\begin{definition}[Information Bonus]
\label{def:ib}
Let $f_{\mathrm{AF2}^*}$ be the stronger AF2 representation for task
$\mathcal{T}$, chosen between single and pair-diagonal Evoformer features, and
let $f_{\mathrm{ESM}}$ be ESM-2 650M embeddings.  For scalar metric $M$ under
protein-level GroupKFold,
\begin{equation}
  \mathrm{IB}(\mathcal{T})
  = M(f_{\mathrm{AF2}^*}, \mathcal{T})
  - M(f_{\mathrm{ESM}}, \mathcal{T}).
\end{equation}
\end{definition}

In practice, we evaluate both AF2 variants on every task and report all
scores; $f_{\mathrm{AF2}^*}$ is selected post-hoc as the variant with the
higher held-out metric.  Because both variants are reported separately
(Tables~\ref{tab:flex},~\ref{tab:allo}), readers can verify the IB for any
AF2 configuration.  This post-hoc selection gives structural representations
a mild upward bias relative to the single ESM-2 baseline; we note that AF2
still yields negative IB on two of four evaluations despite this advantage.

IB$>0$ indicates usable structural signal; IB$<0$ indicates that sequence
embeddings are sufficient or superior.  We use Pearson $r$ for regression and
AUROC for classification.  IB is intentionally a frozen-feature metric: it
measures information available to a downstream system without expensive
fine-tuning, model surgery, or task-specific representation learning.

\section{Experimental Setup}
\label{sec:setup}

\paragraph{Tasks.}
\textbf{Binding affinity} uses PDBbind \citep{liu2017pdbbind} to predict
$-\log_{10}(K_d)$ for 5,680 protein-ligand complexes from mean-pooled protein
features alone (no ligand representation is provided to the probe; see
Section~\ref{sec:results} for interpretive consequences).
\textbf{Flexibility} uses ATLAS \citep{vander2024atlas}, with 268
proteins and 50,426 residues, to predict RMSF by regression and by a balanced
within-protein median classification task.  \textbf{Allostery} uses
AlloSigDB-derived labels \citep{li2022allosigdb}, with 47 proteins, 9,925
residues, and 4.8\% positives, to identify allosteric sites.

\paragraph{Representations and probes.}
AF2 single features are final Evoformer single representations
($\mathbb{R}^{L\times256}$).  AF2 pair-diagonal features use $z_{ii}$
($\mathbb{R}^{L\times128}$), a residue-level view of pairwise geometry.  ESM-2
features are final-layer embeddings from
\texttt{facebook/esm2\_t33\_650M\_UR50D}
($\mathbb{R}^{L\times1280}$).  AF2 is run in single-sequence mode so both
paths receive the same input.  We use ridge regression ($\alpha{=}1.0$) for scalar targets and
class-balanced logistic regression ($C{=}1.0$) for binary targets, with
$\ell_2$-normalized frozen features and fixed seeds.  Regularization
strength is fixed across representations; because features are
unit-normalized, probe capacity depends primarily on signal content rather
than raw dimensionality.

\paragraph{Splits and statistics.}
The primary protocol is 5-fold protein-level GroupKFold: all residues from a
protein are assigned to the same fold.  This is essential for per-residue
tasks because labels are correlated within proteins.  We do not cluster
proteins by sequence identity before splitting; homologous proteins may
therefore appear in different folds, which could modestly inflate all scores
but affects all representations equally.  We report fold means,
permutation baselines, paired fold tests, and Cohen's $d$ where appropriate.

\section{Results}
\label{sec:results}

\begin{figure}[t]
\centering
\colpaperfig{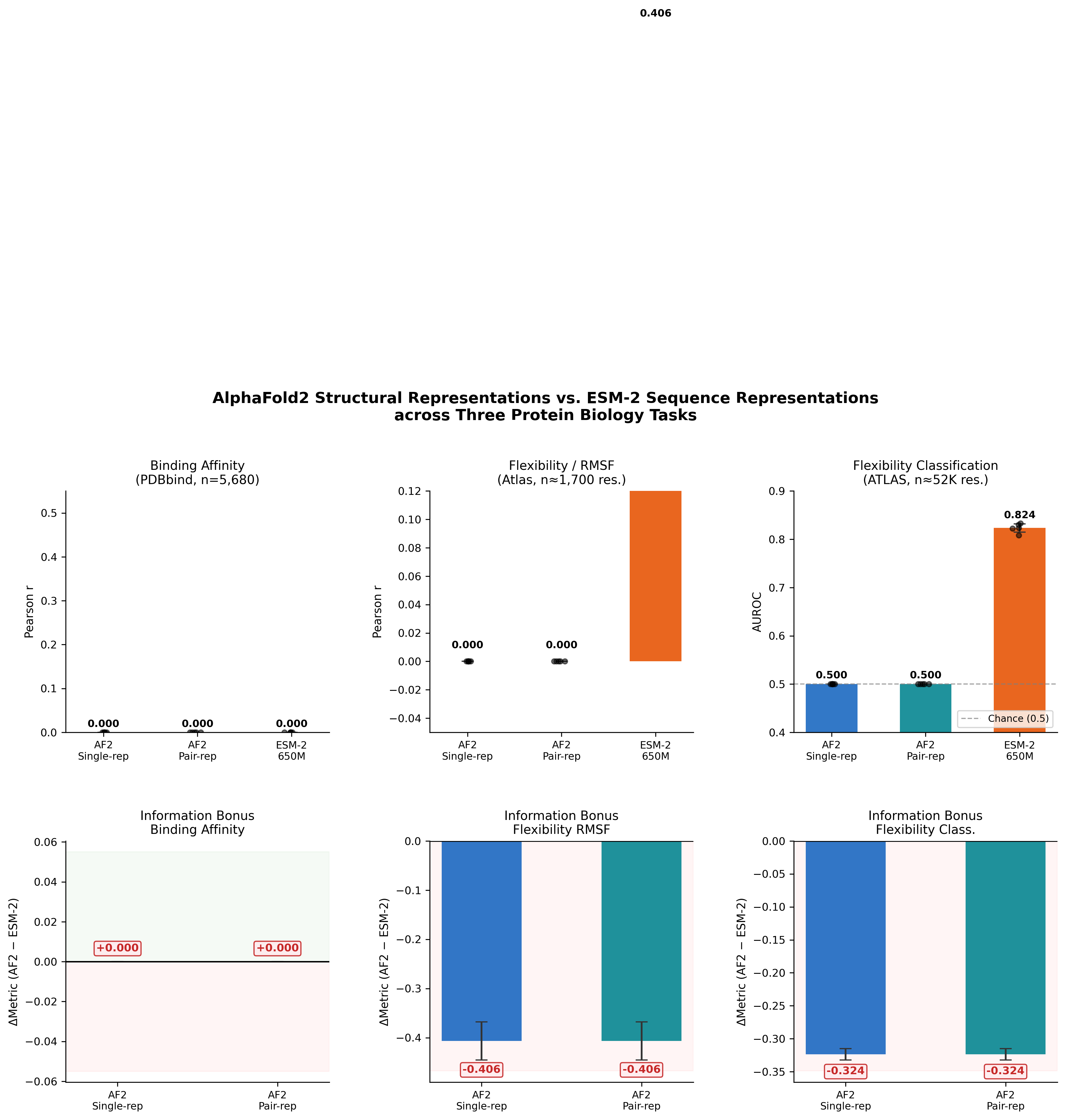}{\columnwidth}{Three-task overview plot.}
\caption{\textbf{Three-task overview.}  ESM-2 wins binding affinity and binary
flexibility, while AF2 is useful when the target depends on long-range
geometry.  Dots show held-out folds.}
\label{fig:main}
\end{figure}

\begin{table}[t]
\caption{\textbf{Main quantitative results.}  Pearson $r$ is used for
regression; AUROC is used for classification.}
\label{tab:main}
\centering
\small
\setlength{\tabcolsep}{4pt}
\renewcommand{\arraystretch}{1.08}
\begin{tabular}{@{}l l c c@{}}
\toprule
Task & Best representation & Score & IB \\
\midrule
Binding affinity & ESM-2 & $.449$ & $-0.141$ \\
RMSF regression & AF2 pair diag. & $.436$ & $+0.030$ \\
Flexibility AUROC & ESM-2 & $.824$ & $-0.060^{***}$ \\
Allostery AUROC & AF2 single & $.548$ & $+0.064$ \\
\bottomrule
\end{tabular}
\vspace{2pt}
{\footnotesize $^{***}p<0.005$ against AF2 pair.  IB is computed from
unrounded fold-level scores; displayed values are rounded independently.}
\end{table}

\subsection{Binding Affinity: Evolutionary Signal Wins}

ESM-2 is the strongest representation for binding affinity, achieving
Pearson $r{=}0.449$ compared with $0.307$ for AF2 single and $0.278$ for AF2
pair-diagonal features.  The structural information bonus is therefore
negative (IB$=-0.141$).  This result is stable across folds.

An important caveat: because the probe receives only protein features (no
ligand representation), the task effectively measures how well each
representation captures protein-level correlates of affinity---pocket
druggability, family-level binding propensity---rather than complex-specific
complementarity.  ESM-2's advantage is therefore best interpreted as stronger
encoding of evolutionary and family-level binding constraints
\citep{lam2024virtualscreen,piao2025bindpred}, not as evidence that sequence
alone can predict ligand-specific $K_d$.  The AF2 features describe the apo
protein, not the ligand-bound complex, which further limits the structural
path.

\begin{figure}[t]
\centering
\colpaperfig{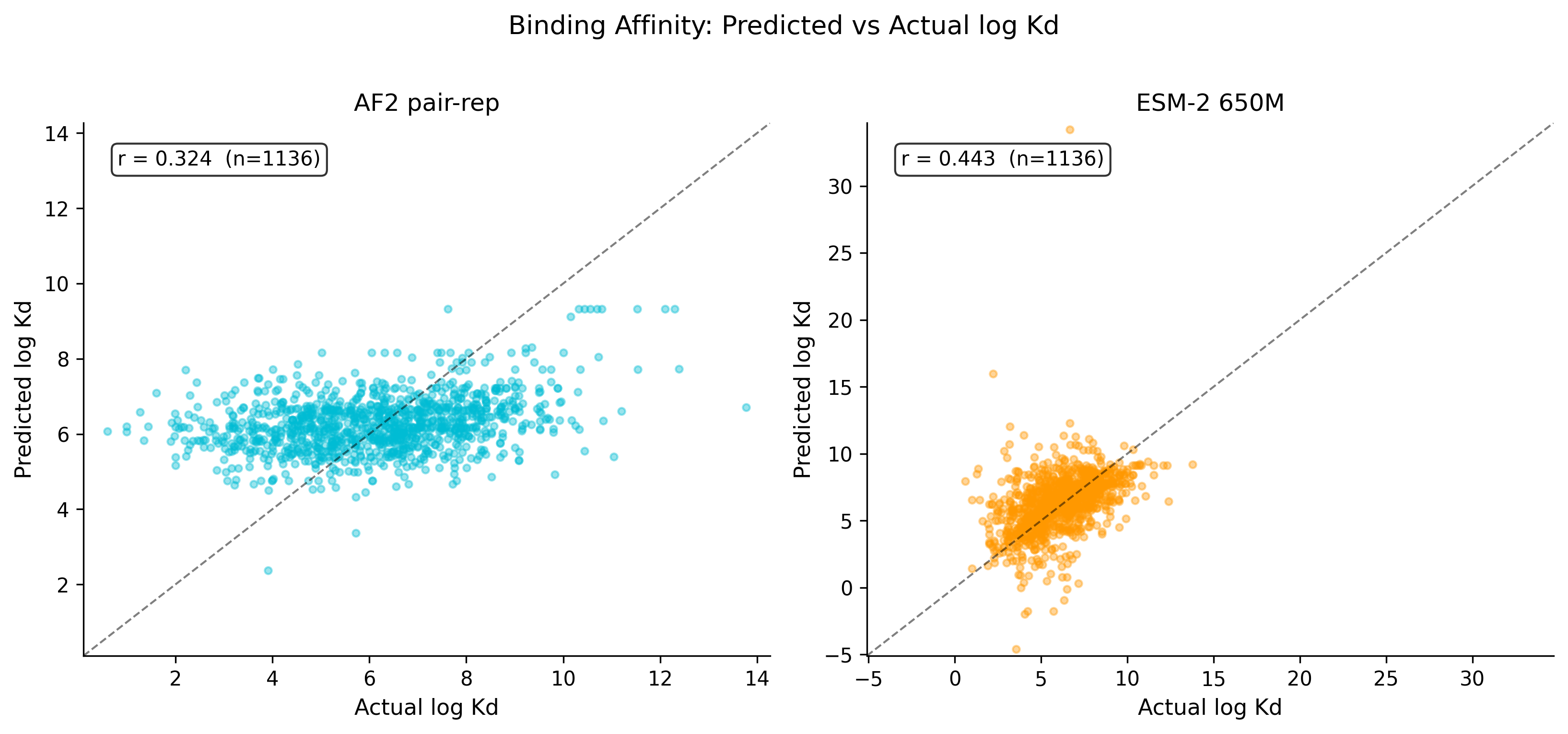}{\columnwidth}{Binding affinity scatter
plot.}
\caption{\textbf{Binding affinity scatter plot.}  ESM-2 yields a higher
predicted-versus-measured correlation ($r=0.443$) than AF2 pair representations
($r=0.324$) on the held-out test fold ($n=1136$).}
\label{fig:binding}
\end{figure}

\subsection{Flexibility: Static Geometry Helps Less Than Sequence Context}

For continuous RMSF regression, AF2 pair-diagonal features are directionally
best ($r{=}0.436$), slightly above ESM-2 ($r{=}0.407$).  This weak positive IB
is consistent with pair-diagonal geometry encoding burial, packing, and
contact density.  However, the advantage is not statistically decisive with
five folds.

For binary flexibility, ESM-2 wins cleanly in every fold, with AUROC $0.824$
versus $0.764$ for AF2 pair and $0.762$ for AF2 single.  The task asks for
relative flexibility within a protein, and sequence models appear to capture
the conserved signatures of disorder and mobility better than static AF2
features.  This is a useful reminder: a task can be physically structural
without requiring an explicitly structural representation.

\begin{table}[t]
\caption{\textbf{Flexibility results on ATLAS.}}
\label{tab:flex}
\centering
\small
\setlength{\tabcolsep}{4pt}
\renewcommand{\arraystretch}{1.08}
\begin{tabular}{@{}l c c c@{}}
\toprule
Target & ESM-2 & AF2 single & AF2 pair \\
\midrule
RMSF $r$ & .407 & .401 & \textbf{.436} \\
Median AUROC & \textbf{.824} & .762 & .764 \\
Permutation & .000/.500 & .000/.500 & .000/.500 \\
\bottomrule
\end{tabular}
\end{table}

\begin{figure}[t]
\centering
\colpaperfig{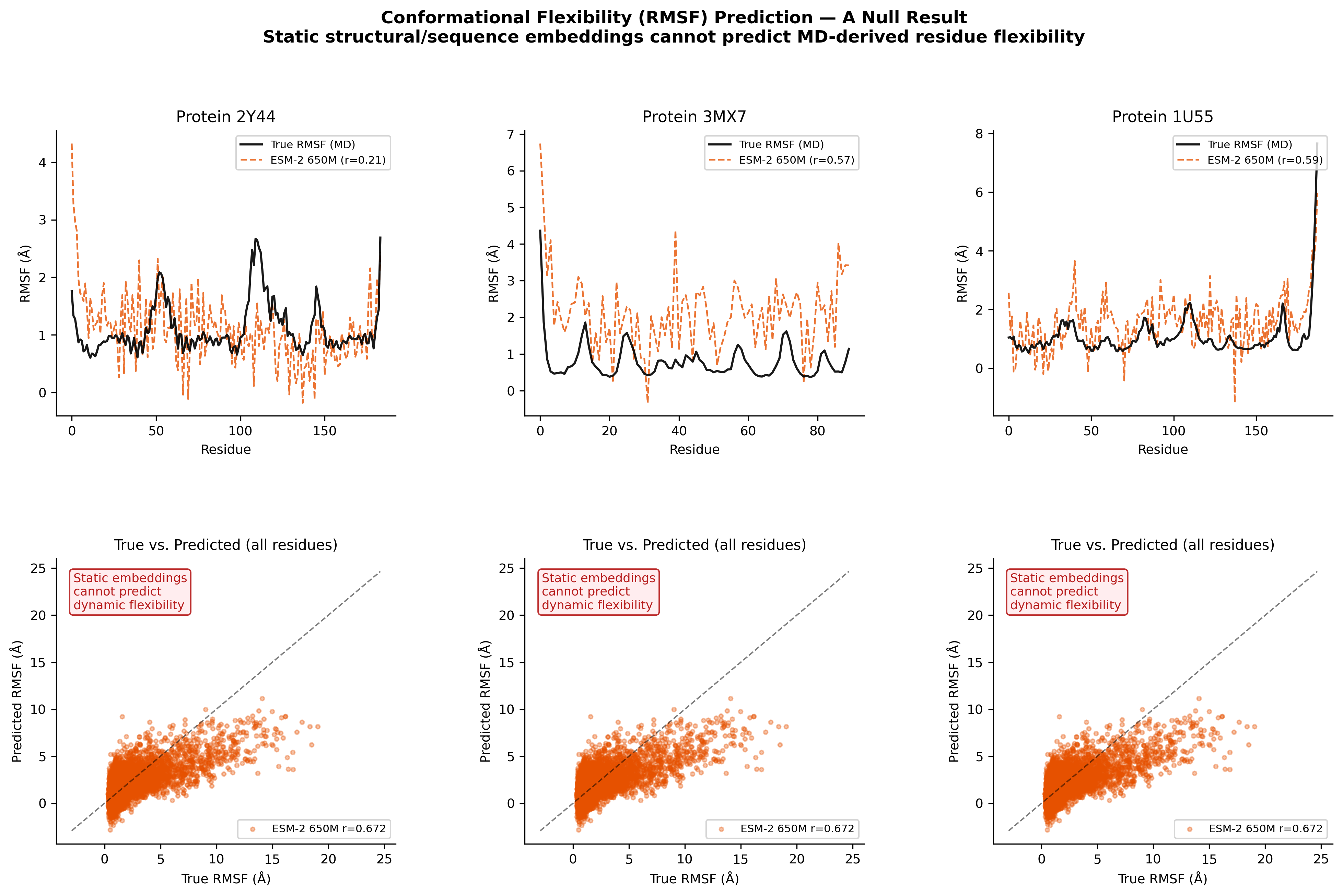}{\columnwidth}{RMSF detail and leakage
comparison.}
\caption{\textbf{RMSF detail and split sensitivity.}  Residue-level splitting
inflates performance by letting train and test residues share protein
identity.}
\label{fig:rmsf}
\end{figure}

\subsection{Residue-Level Splits Inflate RMSF by 27--39\%}

Under naive residue-level 5-fold KFold, ESM-2 reaches $r{=}0.672$ on RMSF.
Under protein-level GroupKFold, it drops to $r{=}0.407$, a 39.4\% reduction.
AF2 pair drops from approximately $0.600$ to $0.436$, a 27.3\% reduction.
Residue-level splitting therefore inflates apparent performance by
27--39\% (relative reduction) depending on the representation, with the
largest effect on ESM-2.

The mechanism is not subtle.  Flexible proteins tend to contain many flexible
residues; rigid proteins tend to contain many rigid residues.  If residues
from one protein are split across train and test sets, the probe can learn
protein-level context instead of transferable residue-level physics.  The
error changes the scientific conclusion: ESM-2 appears dominant under leaky
residue splits, while AF2 pair is directionally best under the correct
protein-level RMSF regression protocol.

\subsection{Allostery: Structure Provides the Missing Signal}

Allostery is the only task with a clearly positive structural information
bonus.  AF2 single reaches AUROC $0.548$, while ESM-2 falls below chance at
$0.485$ and AF2 pair is near chance at $0.497$.  The absolute score is modest
because the dataset is small (47 proteins, ${\sim}$9 per test fold) and
imbalanced, so this result should be interpreted as directional evidence of
structural signal rather than a statistically decisive finding.  Only AF2
single contains above-chance signal.

This matches the biology.  Allosteric residues are defined by their role in a
three-dimensional communication network, not simply by local sequence
patterns.  The Evoformer repeatedly updates residue and pair representations
through global structural context before coordinate prediction.  That geometry
appears to preserve signal about allosteric communication that ESM-2 does not
linearly recover.

\begin{table}[t]
\begin{threeparttable}
\caption{\textbf{Allosteric-site classification} on AlloSigDB.}
\label{tab:allo}
\centering
\small
\setlength{\tabcolsep}{5pt}
\renewcommand{\arraystretch}{1.08}
\begin{tabular}{l c c}
\toprule
Representation & AUROC & IB \\
\midrule
AF2 single & \textbf{.548} & $+0.064$ \\
AF2 pair diag. & .497 & $+0.012$ \\
ESM-2 & .485 & --- \\
Permutation & .500 & --- \\
\bottomrule
\end{tabular}
\begin{tablenotes}\footnotesize
\item AF2 single is the only representation above the chance baseline.
\end{tablenotes}
\end{threeparttable}
\end{table}

\begin{figure}[t]
\centering
\colpaperfig{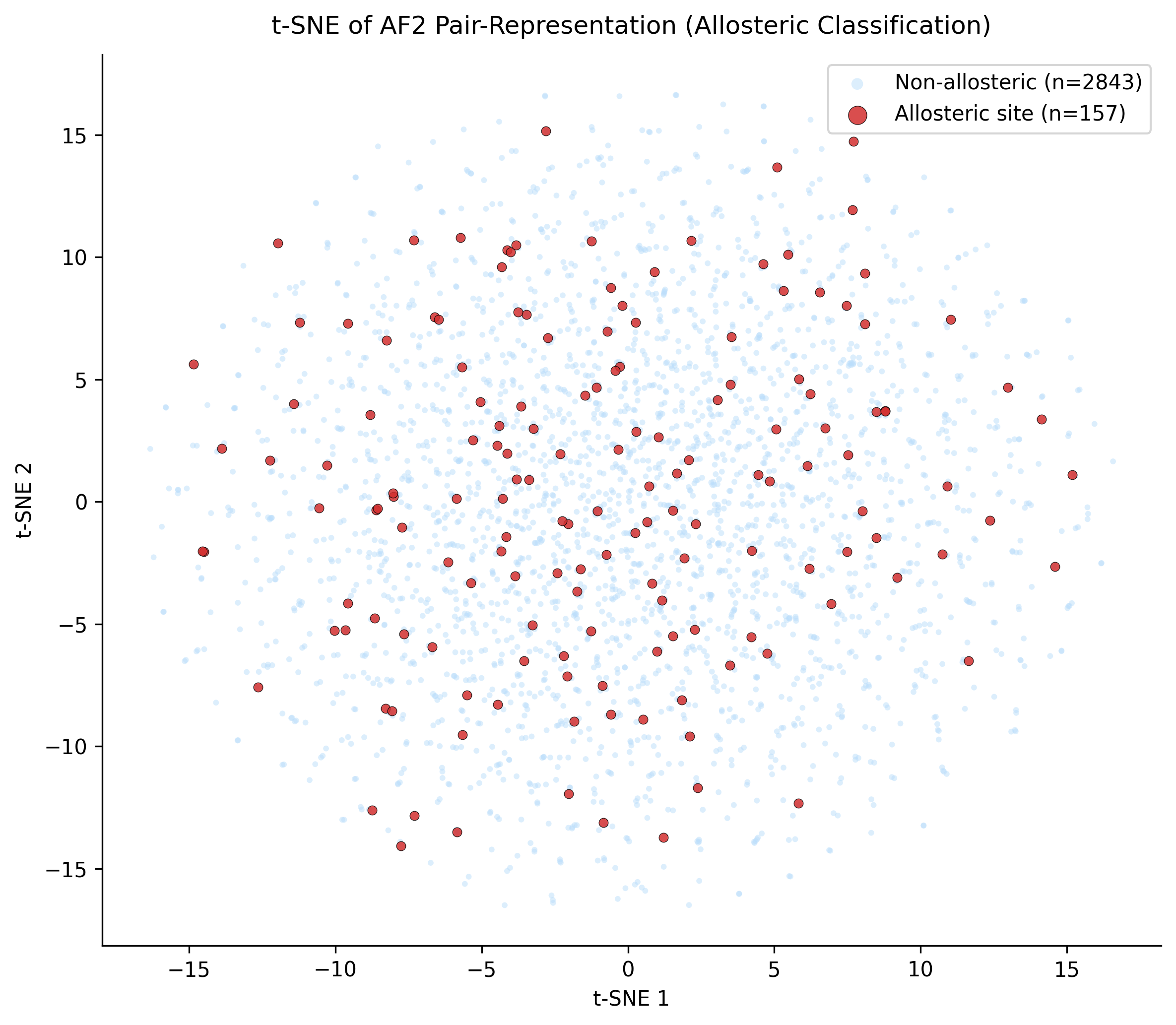}{0.88\columnwidth}{t-SNE allostery plot.}
\caption{\textbf{AF2 representation space for allostery.}  Allosteric
residues show non-random clustering in AF2 single-representation space,
consistent with geometric encoding of allosteric pathways.}
\label{fig:allo}
\end{figure}

\section{Analysis}
\label{sec:analysis}

\paragraph{The boundary is mechanistic, not architectural.}
The same AF2 representation is unnecessary for binding, weakly useful for
continuous flexibility, and uniquely useful for allostery.  This pattern is
exactly what one would expect if sequence models absorb evolutionary
regularities while structural models preserve information about spatial
communication.  IB therefore acts less like a leaderboard score and more like
a diagnostic: it tells an AI scientist system what kind of biological signal a
task requires.

\paragraph{Why sequence often wins.}
Evolution supplies supervision at a scale no structural database can match.
Binding constraints, disorder signatures, functional motifs, and family-level
selection pressures recur across millions of sequences.  ESM-2 can internalize
these regularities without ever seeing explicit coordinates
\citep{lin2023evolutionary,meier2021language}.  When a downstream label is
largely explained by those regularities, AF2's structural inference can add
cost without adding linearly accessible information.

\paragraph{Why allostery is different.}
Allostery depends on residue placement inside a folded contact network.
Sequence co-variation can hint at coupling, but the task asks whether a
particular residue participates in long-range communication.  AF2 single
features, produced after repeated Evoformer updates, carry enough global
geometric context to rise above chance.  This is the positive case for
structural representations: not structure as decoration, but structure as the
missing variable.

\section{Discussion}
\label{sec:discussion}

\paragraph{Decision rule for AI scientist systems.}
Our results suggest starting with ESM-2 when the target is plausibly governed
by evolutionary constraint, family-level function, disorder, or binding-site
chemistry.  We hypothesize that the AF2 cost is justified when the target
explicitly depends on three-dimensional communication---as demonstrated here
for allostery, and plausibly extending to protein-protein interfaces, cryptic
pockets, and domain coupling \citep{cimermancic2016cryptosite}, though these
extensions remain to be tested.  When the mechanism is uncertain, estimating
IB on a small labeled validation set before scaling structural inference
provides a practical safeguard.

\paragraph{Benchmarking implication.}
The leakage result is as important as the representation result.  Per-residue
benchmarks must split by protein, not by residue, whenever labels are
correlated within proteins.  Otherwise, a model can appear to understand local
biophysics while merely interpolating within proteins it has already seen.
For AI-for-science systems that make downstream experimental decisions, this
kind of evaluation error is not a formatting issue; it changes which model is
selected.

\paragraph{Limitations.}
The allostery dataset is small and imbalanced (47 proteins, ${\sim}$9 per
test fold), so AUROC should be interpreted as directional evidence of
representation signal rather than a deployment-ready classifier; fold-level
standard deviations and formal significance tests would strengthen this
result.  The RMSF comparison has limited fold-level statistical power.
Single-sequence AF2 gives the cleanest matched-input comparison but may
underestimate the advantage of full-MSA AF2 on geometry-dependent tasks;
IB measured with full-MSA AF2 could differ substantially.  Regularization
hyperparameters are fixed across all representations rather than tuned
per-representation via nested cross-validation, which could modestly affect
absolute scores.  Finally, linear probes measure accessible information;
nonlinear fine-tuning could improve absolute performance, but would also
change the cost and safety profile of the system.

\section{Conclusion}
\label{sec:conclusion}

We introduced the information bonus, a practical metric for deciding when
single-sequence AlphaFold2 structural representations add usable information
over ESM-2 sequence embeddings.  Across three protein tasks, the answer is
mechanism-dependent.  ESM-2 wins when evolutionary signal is sufficient; in
our experiments, AF2 provided a clear advantage only when long-range geometry
was the missing information.  The same study also shows that residue-level
cross-validation can inflate RMSF performance by 27--39\% and reverse
representation rankings.  For AI scientist systems, structure is powerful,
but it should be treated as a measured scientific choice rather than a default
assumption.

\section*{Broader Impact}

This work improves the reliability of biological AI evaluation by making
representation selection explicit and by identifying a common leakage mode in
per-residue benchmarks.  Better evaluation can reduce unnecessary structural
inference costs and lower the risk of selecting the wrong model for
closed-loop discovery.  We do not anticipate direct dual-use concerns from the
evaluation framework itself.

\bibliography{example_paper}
\bibliographystyle{icml2026}

\end{document}